\documentclass[prb,preprint,amsmath,showpacs]{revtex4}
\usepackage{graphics}

\begin{document}

\input epsf

\title {Thermal conductivity of Al-doped MgB$_2$: impurity scattering and the validity of
the Wiedemann-Franz law}

\author {A.V. Sologubenko\footnote[1]{Current address: \textit{II. Physikalisches Institut, Universit\"{a}t zu K\"{o}ln, 50937 K\"{o}ln, Germany}}, N.D. Zhigadlo,  J. Karpinski, H.R. Ott}
\affiliation{Laboratorium f\"ur Festk\"orperphysik, ETH H\"onggerberg,
CH-8093 Z\"urich, Switzerland}

\date{\today}

\begin{abstract}
We report data on the thermal conductivity $\kappa(T,H)$ along the basal plane of the hexagonal crystal structure of superconducting Mg$_{1-y}$Al$_y$B$_{2}$ with $y=  0.02$ and 0.07 at temperatures between 0.5 and 50~K and in external magnetic fields between 0 and 70~kOe. The substitution of Al for Mg leads to a substantial reduction of the heat transport via electronic quasiparticles. The analysis of the $\kappa(T,H)$ data implies that the Al impurities provoke an enhancement of the intraband scattering rate, almost equal in magnitude for both the $\sigma$- and the $\pi$ band of electronic excitations. This is in contrast with conclusions drawn from analogous data sets for material in which carbon replaces boron and where mainly the intraband scattering rate of the $\sigma$ band is enhanced. Our complete data set, including new results of measurements of the low-temperature thermal conductivity of pure MgB$_2$, confirms the validity of the Wiedemann-Franz law for both pure and doped MgB$_2$.

\end{abstract}
\maketitle

\section{Introduction} 

In multiband superconductors, the onset of superconductivity may introduce gaps of different size on different parts of the Fermi surface. As a consequence, the thermal conductivity of multiband superconductors at $T < T_c$, the critical temperature, and in external magnetic fields $H \leq H_{c2}$, the upper critical field, reveals specific features that  are usually not observed for common type II superconductors. These features were first reported for MgB$_2$ (ref.~\onlinecite{Sologubenko02_KH}) and subsequently also identified for NbSe$_2$ (ref.~\onlinecite{Boaknin03}) and PrOs$_4$Sb$_{12}$ (ref.~\onlinecite{Seyfarth05}). The thermal conductivity $\kappa(H)$ of these materials in the mixed state of superconductivity, measured at constant temperature, exhibits an unusual two-step increase with increasing magnetic field between $H_{c1}$ and $H_{c2}$. The two steps reflect the separate recovery of the heat transport carried by quasiparticles of two electronic energy bands with significantly different energy gaps in the superconducting state.\cite{Kusunose02,Tewordt03_Two} 
In order to observe this type of behaviour of $\kappa_e(H)$, the electronic contribution to the total thermal conductivity, it is important that in the corresponding temperature and magnetic-field ranges, the heat transport carried by phonons and represented by $\kappa_{\rm ph}(H)$, is relatively weak. In addition, also the conditions of weak interband scattering of the quasiparticles and a scattering of quasiparticles in the band developing the smaller gap compatible with a moderately dirty limit situation, need to be fulfilled.\cite{Kusunose02} All these conditions are ideally met for MgB$_2$ at helium temperatures and the two step enhancement of  $\kappa_e(H)$ with increasing field is very pronounced, such that the contribution of each of the two types of band to  $\kappa_e$ can be identified. Hence this allows, in principle, to investigate the influence of different perturbations on the heat transport carried by different energy carriers separately. Indeed, results of measurements of the thermal conductivity of Mg(B$_{1-x}$C$_x$)$_2$ (Ref.~\onlinecite{Sologubenko05}) revealed that low concentrations of C replacing B reduce the step-like enhancement of $\kappa(H)$ near $H_{c2}$ much more than the analogous feature of $\kappa(H)$ which is observed at much lower fields. This result suggests that the carbon related defects scatter quasiparticles of the band with the large gap much more than quasiparticles excited in the band with the smaller gap. This observation agrees with the expectation that C substitutions for B mostly enhance the scattering between states in the $\sigma$ band which is predominantly formed by boron $s p_x p_y$ orbitals. In this work we describe and discuss results of analogous measurements of $\kappa(H,T)$ of  Mg$_{1-y}$Al$_y$B$_{2}$. The investigations of single crystalline material with $y = 0.02$ and 0.07 reveal that Al replacing Mg at the few at-\% level reduces the heat transport of quasiparticles in both the  $\sigma$- and the $\pi$ band in a more or less equal manner.

Aluminium replacing Mg in MgB$_2$ is expected to influence the superconductivity of the pure binary compound mainly via two effects.\cite{Erwin03,Kortus05}
First, Al doping causes a reduction of hole-type carriers, thus reducing the corresponding density of electronic states (DOS) and, likewise, the size of both gaps $\Delta_{\sigma}$ and $\Delta_{\pi}$, thereby accounting for the reduction of T$_c$. The second consequence is caused by out of plane distortions created by Al occupying Mg sites which enhance the interband scattering thus leading to a relative increase of the gap in the $\pi$ band. Indeed, results of various experimental studies confirm this view and suggest that for $y \lesssim 0.1$, the replacement of Mg by Al leads to a gradual reduction of $\Delta_{\sigma}$ without substantial changes in the value of $\Delta_{\pi}$.\cite{Putti03_Al,Gonnelli06,Putti05,Cooley05}
For $y > 0.1$ the situation is more controversial but most experimental results are compatible with the reduction of both gaps upon increasing $y$.\cite{Putti05,Gonnelli06,Klein06} Because choosing $y > 0.1$ often leads to inhomogeneities in the Al distribution, such as the segregation of samples into Al-rich and Al-poor regions,\cite{Karpinski05,Gonnelli06} we restricted our study to material with $y < 0.1$, for which the influence of Al is expected to be as described above.

\section{Experiment}

The preparation and characterization of the  Mg$_{1-y}$Al$_y$B$_2$ single crystals that were used in this study is described in detail in ref.~\onlinecite{Karpinski05}. The synthesis under high pressure allowed the production of single-phase, single-crystalline material up to Al concentrations  $y \sim 0.1$. For our measurements we selected two samples, below denoted as Al2 and Al7, with Al contents, as determined by EDX, of $y = 0.02$ and  0.07, respectively. Since we also intended to confirm or refute our earlier conclusion of the non validity of the Wiedemann-Franz law for MgB$_2$ at helium temperatures (ref.~\onlinecite{Sologubenko02_KH}), we repeated the measurement of the thermal conductivity of binary MgB$_2$, using two newly synthesized single crystals, denoted as MB1 and MB2 below. All samples were bars with the approximate shape of a prism and typical dimensions of $0.9 \times 0.3 \times 0.06$~mm$^3$. The thermal conductivity was measured using the same standard uniaxial heat-flow method that we applied in our previous experiments on pure and C-doped MgB$_2$.\cite{Sologubenko02_KH} 
In this set-up, one end of the sample is attached to a copper body serving as a heat sink and at the opposite end, a miniature RuO$_2$ resistor, serving as the heater, is attached. The temperature gradient along the sample was monitored employing either RuO$_2$ resistor thermometers below 4~K or chromel/Au+0.07\%Fe thermocouples above 4 K. 
These temperature sensors were each attached to a  25~$\mu$m diameter gold wire and the wires were glued to the crystals with silver-filled epoxy, approximately 300 to 400 ~$\mu$m apart along the direction of the heat flow. The temperature difference along the sample was chosen to be of the order of 3\% of the average sample temperature. For the subsequent measurements of the dc electrical resistivity $\rho$, a four contact configuration was used. In order to minimize the uncertainty of the geometrical factor in the comparison of the electrical and thermal conductivities, the described thermometer contacts were also used as voltage contacts for the resistivity measurements. Nevertheless, the possible systematic errors in the separate evaluation of $\kappa$ or $\rho$ values caused by the uncertainty of the sample geometry were as large as $\pm$20\%, whereas the random errors for  $\kappa$ and $\rho$ were only $\pm$1\% and $\pm$0.1\%, respectively.

\section{Results}

The electrical resistivity was measured between 2 and 300~K in zero magnetic field and $H = 50$~kOe, oriented along the crystallographic $c$ direction. For both new specimens of pure MgB$_2$, MB1 and MB2, $\rho(T)$ is very similar to the data  that we obtained in earlier work (ref.~\onlinecite{Sologubenko02_KH}) for sample MB0.  Likewise the values of T$_c$, the transition widths and the residual resistance ratios turned out to adopt similar values as were established in our previous work. Note, however, that the absolute resistivities of the new samples just above T$_c$ are somewhat lower than before (see Table~\ref{Table1}). In the temperature range between T$_c$ and approximately 150~K, $\rho(T)$ may very well be represented by 
$\rho(T) = \rho_0 + a T^{3}$. 
The most obvious variation of $\rho$ upon replacing small amounts of Mg by Al is an expected increase of the residual resistivity $\rho_0$, already reached at 40~K, which is caused by an enhanced scattering rate due to the Al defects in the Mg sublattice. The temperature dependent component due to the scattering of electrons by phonons changes only very little.

\begin{table}[tb]
\
  \centering
  \caption{
  Samples of Mg$_{1-y}$Al$_y$B$_2$ studied in this work. Sample MB0, listed for comparison, was studied earlier (see ref.~\onlinecite{Sologubenko02_KH}).  
  } %
  \label{Table1} %
\begin{tabular}{ccccc}
        \hline \hline
        & $y$ & $T_c$ (K) & $\rho$(40~K)  & $\rho$(300~K)/$\rho$(40~K) \\ 
        &  & from $\rho(T)$  &  ($\mu\Omega$~cm) &  \\ 
        \hline
    MB0 & 0  & $38.1 \pm 0.1$ & 2.05 & 6.80 \\
    MB1 &  0 & $38.3 \pm 0.1$  &  0.71 & 7.49 \\
    MB2 &  0 & $38.2 \pm 0.1$  &  1.20 & 7.11 \\
    Al2  & 0.017 & $35.9 \pm 0.15$ & 2.36 & 3.90 \\    
    Al7  & 0.073 & $33.0 \pm 0.2$ &   4.13 & 2.70 \\    
    \hline \hline
\end{tabular}
\end{table}

\begin{figure}[t]
 \begin{center}
  \leavevmode
  \epsfxsize=0.8\columnwidth \epsfbox {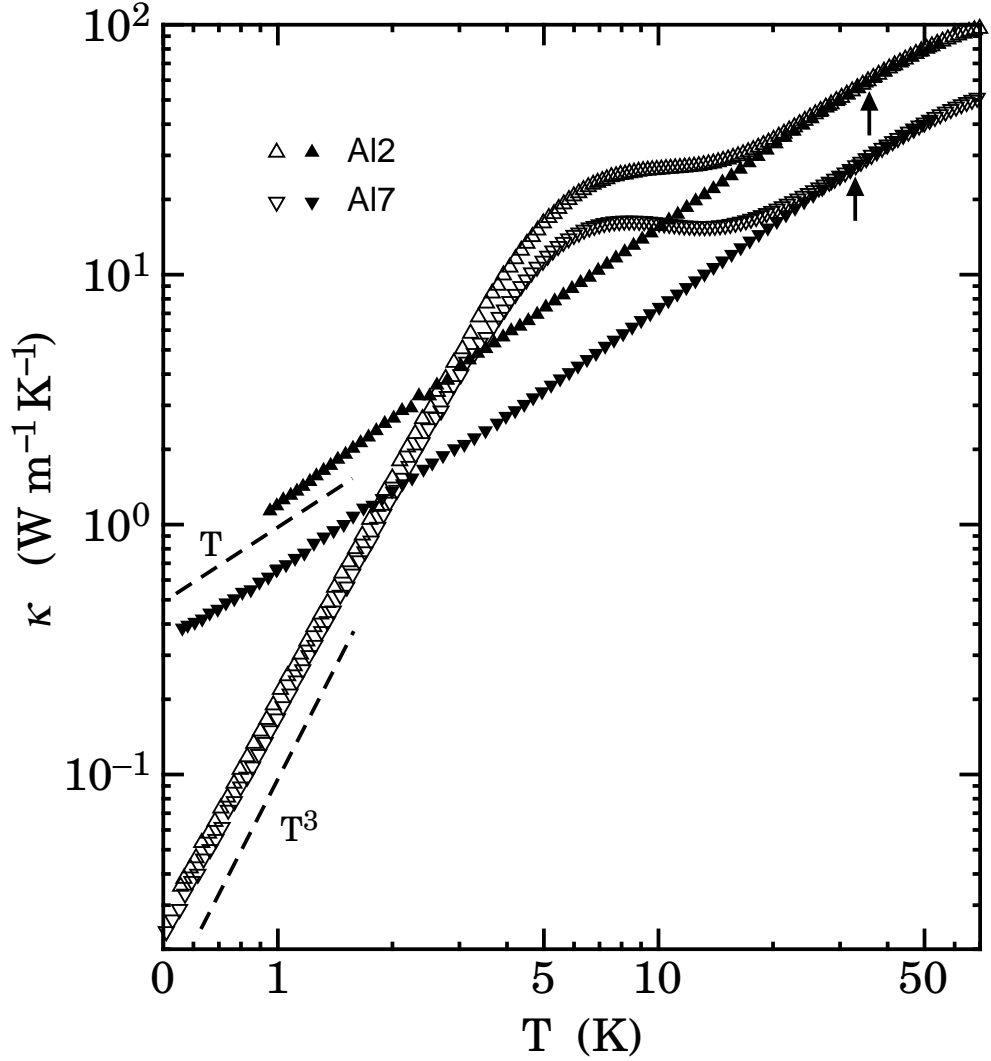}
   \caption{
   Thermal conductivity vs temperature in the ab plane of single-crystalline Mg$_{1-y}$Al$_y$B$_2$ ($y = 0.02$ and 0.07) in  zero magnetic field (open symbols) and $H \parallel c = 50 {\rm ~kOe}$ (solid symbols). The arrows indicate the critical temperatures in zero magnetic field.}
\label{KT}
\end{center}
\end{figure}

In Fig.~\ref{KT}  we display the thermal conductivity of Mg$_{1-y}$Al$_y$B$_{2}$ ($y = 0.02$ and 0.07) as a function of temperature in zero magnetic field and for $H = 50 {\rm ~kOe}$, with $H$ oriented parallel to the $c$ axis. In zero magnetic field, the absolute difference in magnitude of $\kappa$ is more pronounced at elevated temperatures but with decreasing temperature to much below $T_c$ this difference is reduced significantly and is only marginal at temperatures below 3~K. Below 3~K, the temperature dependence of $\kappa$ is fairly well represented by  $\kappa(T) \propto T^3$. In the field induced normal state, the relative difference in the $\kappa$ values persists to temperatures below 1~K where $\kappa(T)$ approaches a variation linear in temperature.

\begin{figure}[t]
 \begin{center}
  \leavevmode
  \epsfxsize=1\columnwidth \epsfbox {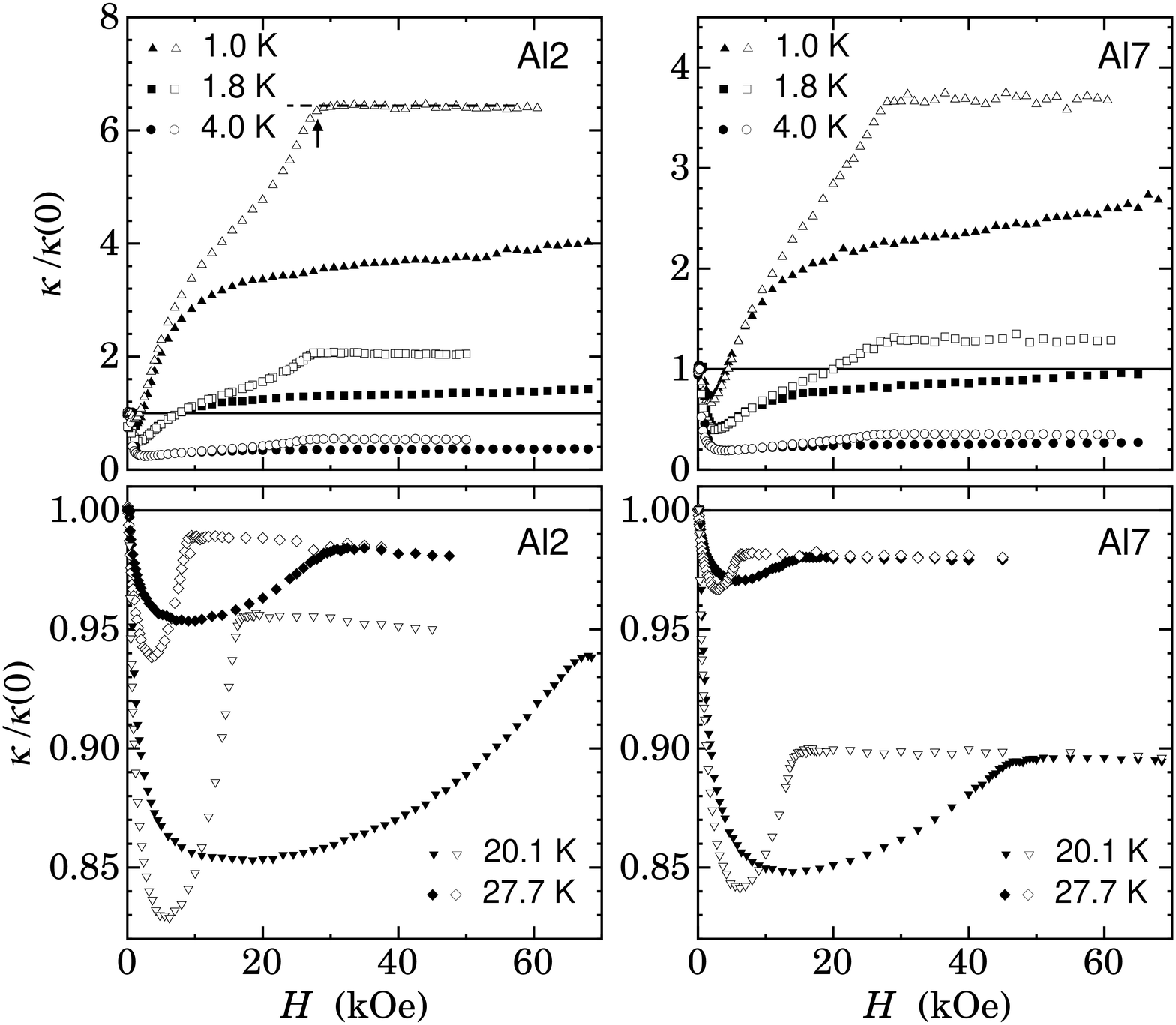}
   \caption{
   Thermal conductivity in the basal plane of Mg$_{1-y}$Al$_y$B$_{2}$ ($y= 0.02, 0.07$) vs $H$ at several fixed temperatures.  The closed and open symbols correspond to the field directions perpendicular and parallel to the $c$ axis, respectively. The arrow in the upper left panel demonstrates how $H_{c2}$ for a given field orientation and at a given temperature was established.
  }
\label{KH}
\end{center}
\end{figure}

\begin{figure}[t]
 \begin{center}
  \leavevmode
  \epsfxsize=0.8\columnwidth \epsfbox {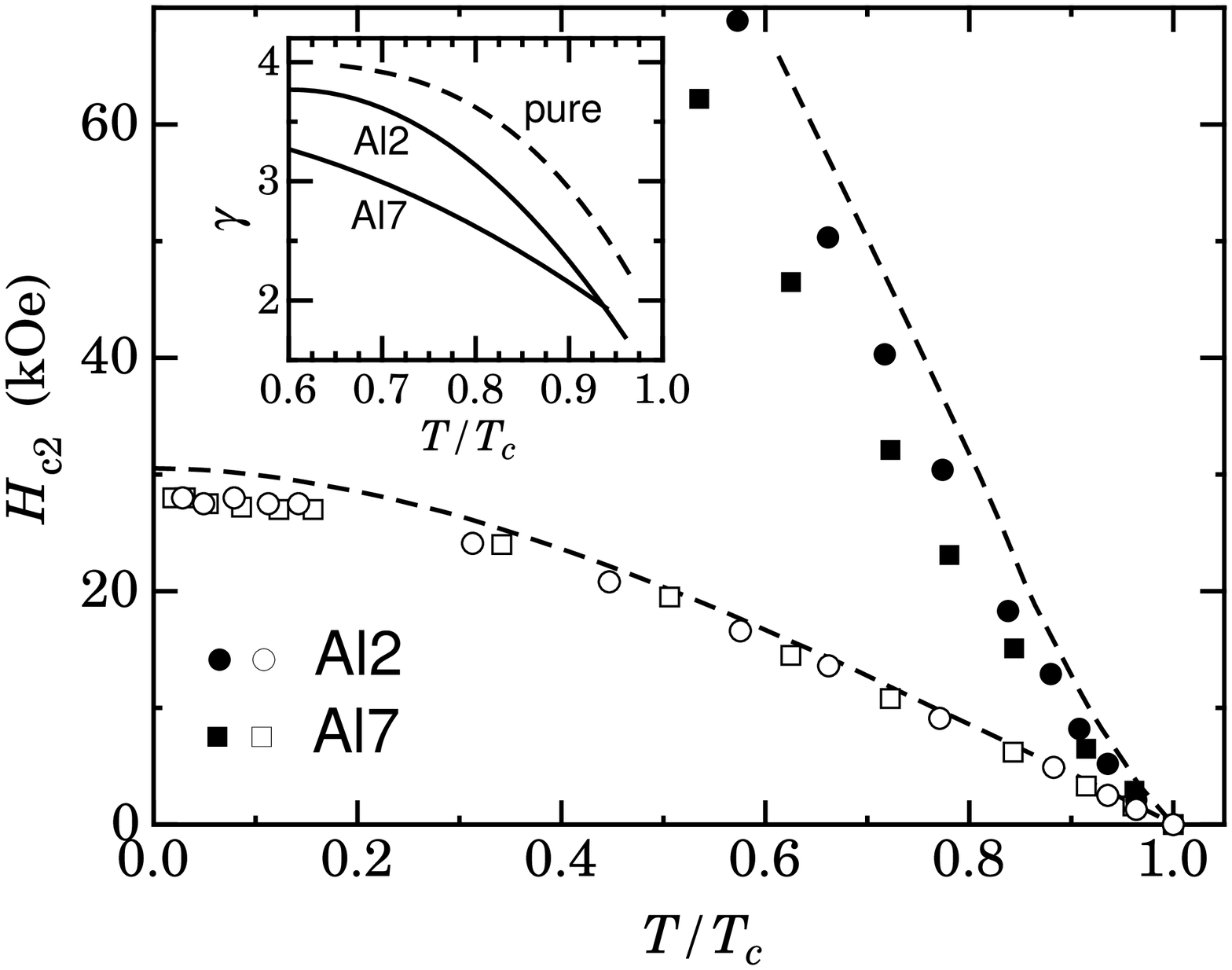}
   \caption{
   The upper critical fields $H_{c2}(T/T_c)$. The closed and open symbols correspond to the field directions perpendicular and parallel to the $c$ axis, respectively. For comparison, our earlier results for pure MgB$_2$ (Ref.~\onlinecite{Sologubenko02_Hc2}) are shown by the dashed lines. The inset shows the anisotropies of the upper critical fields $\gamma = H_{c2}^a/ H_{c2}^c$.
  }
\label{Hc2}
\end{center}
\end{figure}

In Fig.~\ref{KH} we display typical curves for $\kappa(H)$, obtained for the two investigated samples at various constant temperatures. Below $H_{c1}$ and above  $H_{c2}$, the thermal conductivity varies only weakly with field. However, in the mixed state, i.e., between $H_{c1}$ and  $H_{c2}$, $\kappa(H)$ exhibits a substantial field dependence, as expected. While $H_{c1}$ cannot reliably be determined from our data sets because the $\kappa(H)$ curves are irreversible in the vicinity of $H_{c1}$, the upper critical field can be identified quite accurately from the abrupt change of slope of $\kappa(H)$, as is indicated by the arrow in the upper left panel of Fig.~\ref{Hc2}. The resulting curves of $H_{c2}(T)$ are shown in Fig.~\ref{Hc2}. We emphasize, as in our previous work, that $H_{c2}(T)$ obtained in this manner is a reliable measure for the temperature dependence of the upper critical field of the bulk of the sample, insensitive to surface superconductivity or minority-phase inclusions (ref.~\onlinecite{Sologubenko02_Hc2}). The observed trend of the alteration of $H_{c2}(T)$, its values and anisotropies (see inset of Fig.~\ref{Hc2}) upon Al concentration, is consistent with the results presented in ref.~\onlinecite{Karpinski05}. 

\section{Analysis and discussion}

\subsection{Multiband feature of thermal transport}\label{s_KH}

In MgB$_2$ the total heat transport is carried by itinerant lattice and electronic quasiparticle excitations and  $\kappa$ is given by the sum
\begin{equation}\label{eKtot}
\kappa_{\rm tot} = \kappa_{\rm ph} + \kappa_e
\end{equation}
with $\kappa_e = \kappa_{e,\sigma} + \kappa_{e,\pi}$, where the last two terms capture the heat transport by quasiparticles of the $\sigma$- and the $\pi$ band, respectively. In principle, fitting the zero-field $\kappa(T)$ data in the temperature range $T \leq T_c$ by invoking eq.~(\ref{eKtot}) and the appropriate theoretical expressions for the individual terms, estimates of the corresponding gaps in the quasiparticle excitation spectra can be made. The theory of Bardeen, Rickayzen and Tewordt (ref.~\onlinecite{Bardeen59}) has successfully been used to evaluate the energy gaps of conventional, single-gap superconductors from thermal conductivity data. For multi-gap superconductors, more than one channel of quasiparticle heat transport in different energy bands need to be considered (ref.~\onlinecite{Tewordt02}) and therefore, the number of fit parameters increases according to the increasing number of bands involved. If the heat transport carried by phonons is not negligible, the number of independent parameters increases even more. In particular, also the scattering of phonons by quasiparticles has to be taken into account. In our case, the upturn of $\kappa(T)$ in zero magnetic field just below $T_c$ indicates that the phonon contribution to the thermal conductivity in the normal state, $\kappa_{\rm ph}^n$, is not small in comparison with the normal-state component $\kappa_e^n$. Thus $\kappa_{\rm ph}$ is not negligible at $T \lesssim T_c$ and any fitting procedure in the manner indicated above seems unreasonable. Nevertheless, in ref.~\onlinecite{Putti03} an attempt of such an analysis has been made for data obtained for polycrystalline MgB$_2$ where the phonon contribution to the total thermal conductivity is claimed to be negligibly small.

Information about details of quasiparticle scattering can be obtained from the analysis of the magnetic field dependence $\kappa(H)$ of the thermal conductivity, measured in the mixed state of type II superconductors at constant temperature. Usually, upon increasing $H$ above $H_{c1}$, $\kappa(H)$ first decreases because of an additional scattering of both phonons and quasiparticles by vortices. With still increasing $H$, the increasing density of quasiparticles in normal vortex cores enhances $\kappa_e$. Also for MgB$_2$ the thermal conductivity in the mixed state is dictated by such processes but the resulting features of  $\kappa(H)$ are clearly different from those observed for common superconductors. 
First, $\kappa_e$ increases very rapidly in rather low fields, such that a considerable part of the normal-state thermal conductivity is recovered already when $H \ll H_{c2}$. This is distinctly different from observations in common type II superconductors where, depending on the sample purity, $\kappa_e(H)$ either grows significantly and rapidly only if $H$ approaches $H_{c2}$ (pure limit) or reaches its normal state value in an approximately linear fashion, i.e., $\kappa_e \propto H/H_{c2}$ (dirty limit).\cite{Vinen71} 
Second, when measured along different crystallographic directions and taking into account the anisotropy of the upper critical field, the low-field increase of $\kappa(H)$ does not scale as $H/H_{c2}$ as is the case for common type II superconductors. 
For MgB$_2$, these anomalous features find a natural explanation by considering the anisotropy of the gap in the quasiparticle excitation spectrum, i.e., the different gap values for the $\sigma$- and the $\pi$ band, respectively. It turns out that the smaller gap $\Delta_\pi$ is quenched already at $H^* < H_{c2}$, where the critical field $H^*$ is practically independent of the orientation of $H$, reflecting the three-dimensional nature of the $\pi$ band. In this scenario, in the intermediate field regime $H \approx H^* < H_{c2}$, the contribution of the quasiparticles created in the $\pi$ band is close to that in the normal state whereas in the $\sigma$ band, the large gap prevents the excitation of quasiparticles until $H$ approaches $H_{c2}$ and hence, $\kappa_{e,\sigma}$ grows substantially only when $H$ is close to $H_{c2}$. In shape this is similar to what is observed for common, moderately pure type II superconductors. Considering that at a given temperature, $H_{c2}^{ab}$ is significantly larger than $H_{c2}^c$, the ratio $\kappa_e(H \perp c \approx H^*)/\kappa_e(H\parallel c \geq H_{c2})$ provides a rough estimate of the ratio $\kappa^n_{e,\pi}/\kappa^n_{e}$.  

\begin{figure}[t]
 \begin{center}
  \leavevmode
  \epsfxsize=0.8\columnwidth \epsfbox {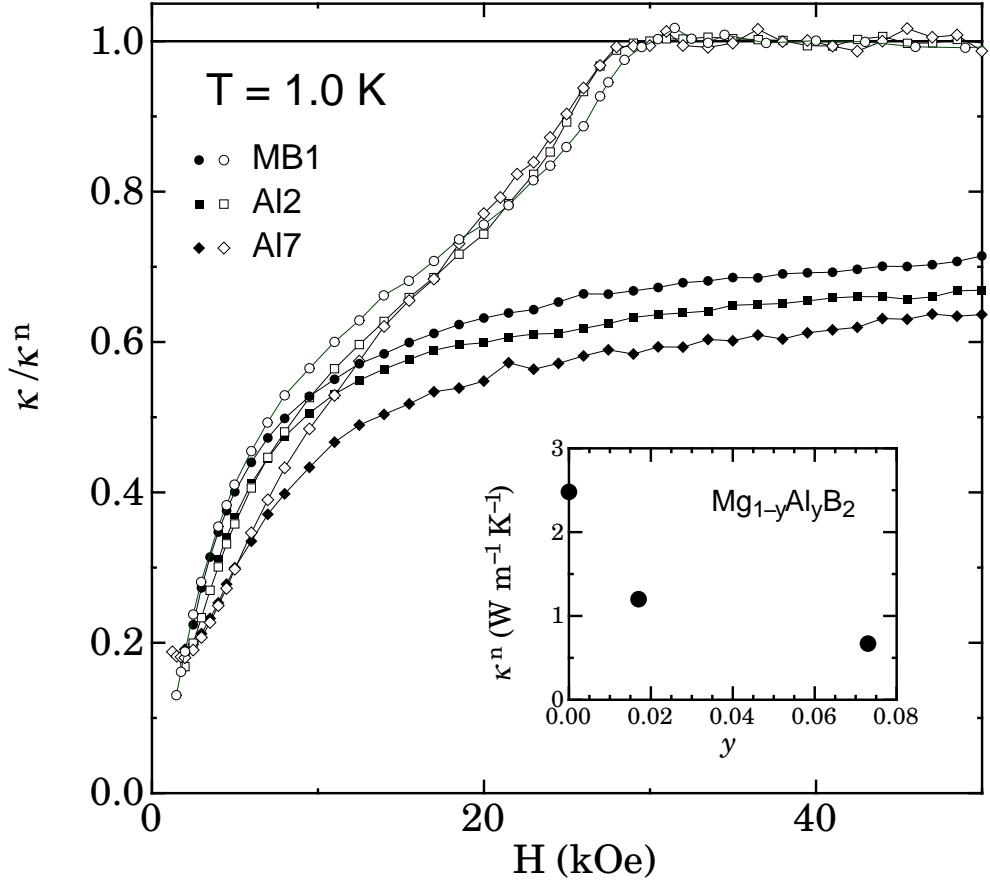}
   \caption{
   Reversible thermal conductivity $\kappa$ normalized to its normal-state value $\kappa^n$ vs. $H$ at 1.0~K for pure and Al-doped MgB$_2$. The closed and open  symbols correspond to the field directions perpendicular and parallel to the $c-$axis, respectively. The inset shows doping dependence of $\kappa^n$ at 1.0K. }
\label{KH1K}
\end{center}
\end{figure}
In Fig.~\ref{KH1K} we compare $\kappa(H)$, measured at $T = 1.0$~K, for pure and Al-doped MgB$_2$; only data points where $\kappa(H)$ is reversible are included in the plot. 
In this temperature range, $\kappa_{\rm ph} \ll \kappa_e^n$ and therefore, the field dependence of $\kappa$ mostly reflects that of the thermal conductivity due to quasiparticles. The diagram clearly indicates that Al substitutions for Mg reduce the ratio $\kappa^n_{e,\pi}/\kappa^n_e$ and thus $\kappa^n_{e,\pi}/\kappa^n_{e,\sigma}$ but the latter ratio changes much less than the total electronic thermal conductivity in the normal state $\kappa^n_e$ (see inset of Fig.~\ref{KH1K}). For the ratios $\kappa^n_{e,\pi}/\kappa^n_{e,\sigma}$ at 1~K, we estimate the values of 1.9, 1.6 and 1.2 for the samples MB1, Al2 and Al7, respectively. This suggests that although the introduction of Al for B reduces the thermal conductivity of quasiparticles slightly more strongly in the $\pi$ band, the overall effect is nearly the same in both channels. This is in stark contrast to the observations made in the case of Mg(B$_{1-x}$C$_x$)$_2$ where a 3\% substitution of B by C reduces the quasiparticle thermal conduction of the $\sigma$ band to a negligible level in comparison to the $\pi$ band contribution.\cite{Sologubenko05}   

The general equations capturing the thermal conductivity due to quasiparticles can be simplified to the form
\begin{equation}
\label{eKeQP}
\kappa_e  = \frac{1}{3} C_{Ve} v_F l_e,
\end{equation}
where $C_{Ve}$ is the electronic specific heat, $v_F$ the Fermi velocity, and $l_e$ the electronic mean free path. In the mixed state the mean free path is given by $l_e = v_F/(\gamma + \gamma_A)$, where $\gamma$ is the scattering  rate due to impurities and $\gamma_A$ is the scattering rate due to Andreev scattering of qu¹asiparticles at vortices.\cite{Tewordt02} 
In this scenario, only $\gamma$ is affected by Al impurities. For MgB$_2$ this approach has to be modified to
\begin{equation}
\label{eKeQPsp }
\kappa_{e,i}  = \frac{1}{3} C_{Ve,i} v_{F,i} l_{e,i},
\end{equation}
with $i = \pi, \sigma$. It has been established that Al substitutions up to $y = 0.1$ change neither $C_{Ve}$ nor $v_{F,\pi}$ in the $ab$ plane in any significant way.\cite{Klein06}  With an Al content $y = 0.1$, $v_{F,\sigma}$ in the $ab$ plane is reduced only by 10 to 15\% with respect to its value in the pure compound.\cite{Putti05,Klein06} This suggests that the reduction of the quasiparticle mean free path $l_e$ is responsible for the substantial reduction of $\kappa_e$ shown in the inset of Fig.~\ref{KH1K}. 
If we assume that both the specific heat and the Fermi velocities remain indeed unchanged, our data imply that a 7.3 at.\% substitution of Mg by Al leads to reductions of $l_{e,\sigma}$ and $l_{e,\pi}$ by factors of 2.7 and 4.5, respectively. A de Haas-van Alphen study of Mg$_{1-y}$Al$_y$B$_2$ (ref.~\onlinecite{Carrington05}) indicates that for $y \approx 0.075$, the mean free path on the $\sigma$ sheet of the Fermi surface is approximately a factor of 2 shorter than for pure MgB$_2$ and for the $\pi$ sheet, the reduction of $l_{e,\pi}$ is estimated to be of the order of 3 or slightly larger. These results are obviously in very good agreement with our conclusions.

\subsection{Relations between thermal and electrical conductivity}\label{s_WFL}

The relaxation rates which govern the electronic charge- and heat currents in metals, respectively, are in general not identical. However, some of the involved scattering processes are, under certain conditions, equally efficient in relaxing both electric and thermal currents. In these cases, the electrical resistivity $\rho$ and the electronic thermal conductivity $\kappa_e$ are related via the law of Wiedemann and Franz (WFL), such that
\begin{equation}\label{eWFL}
\kappa_{e}(T) = L_{0} T / \rho(T),
\end{equation}
where $L_{0}=2.45\times 10^{-8}$ W~$\Omega$ K$^{-2}$ is the Lorenz number.
This relation is expected to be valid for any metal whose electronic subsystem can be described as a Fermi liquid. Elastic scattering of electrons by static lattice defects is one of the mechanisms which are compatible with the validity of the WFL. If this type of scattering dominates all other processes, the residual resistivity at low temperatures adopts a constant value and, in the same temperature regime, $\kappa_e$ varies linearly with temperature. The experimental Lorenz number $L(T) = \kappa(T) \rho(T) / T$, with $\kappa(T)$ as the directly measured thermal conductivity, is a constant and equal to $L_0$ only if the phonon thermal conductivity is negligible, i.e., $\kappa_{\rm ph} \ll \kappa_e$ and hence $\kappa_e \approx \kappa$. Inelastic scattering processes always lead to $L(T)<L_0$ and sizable heat transport by phonons provides and excess heat conduction and hence  $L(T)>L_0$.
 
In the field induced normal state, all samples of MgB$_2$ exhibit a temperature independent resistivity below 40 K, suggesting that defect scattering of quasiparticles dominates the low-temperature electronic transport. In passing we note that even in external magnetic fields, oriented along the $c$ axis and of the order of 50~kOe, i.e., exceeding the corresponding bulk value of $H_{c2}$ at $T = 0$~K for all investigated materials, the onset of a superconducting transition is observed at very low temperatures. Most likely this  transition is related to properties of the sample surface and hardly affects the predominantly probed heat transport in the bulk of the sample. In order to check the validity of the WFL, we thus used $\rho=\rho_0 \approx \rho(40 {\rm ~K})$, presented in Table~\ref{Table1}. 

\begin{figure}[t]
 \begin{center}
  \leavevmode
  \epsfxsize=0.8\columnwidth \epsfbox {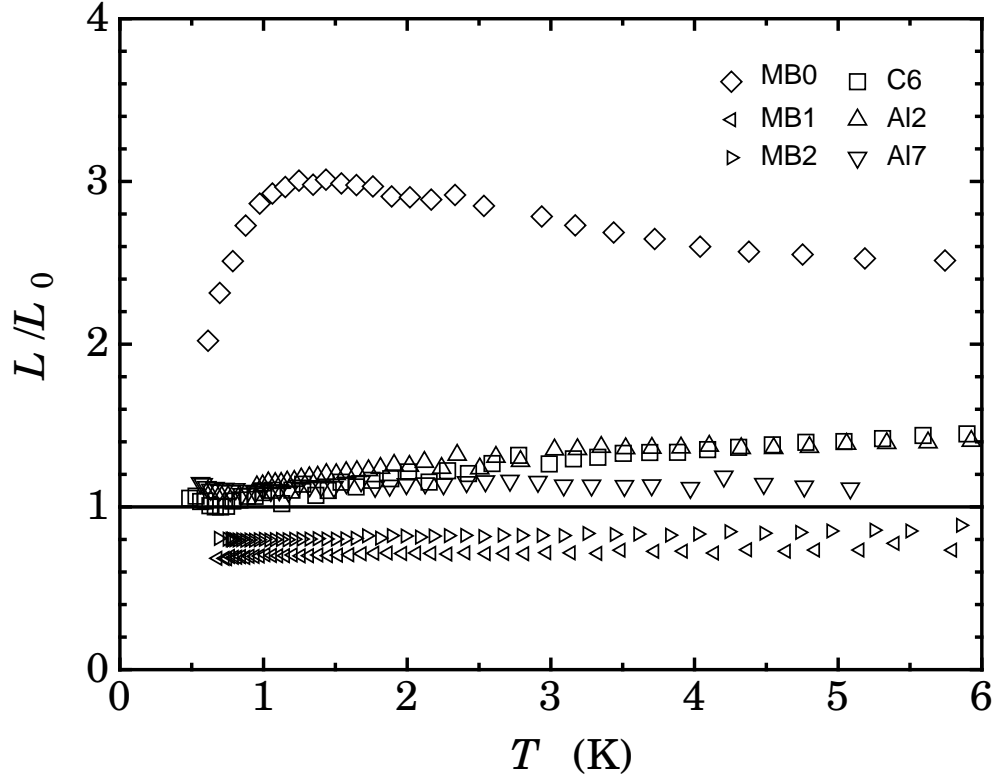}
   \caption{Normalized Lorenz number $L(T)/L_0$ for three samples of pure MgB$_2$ and for C- and Al-doped  MgB$_2$. C6 denotes Mg(B$_{0.94}$C$_{0.06}$)$_2$ (data from Ref.~\onlinecite{Sologubenko05}).}
\label{LL0}
\end{center}
\end{figure}

In part of our earlier work (ref.~\onlinecite{Sologubenko02_KH}) we encountered evidence that for pure MgB$_2$, $L(T)/L_0 > 1$ at low temperatures which could not simply be attributed to a substantial phonon contribution to $\kappa(T)$, at least not below $T = 6 {\rm ~K}$. Our most recent experiments on Mg(B$_{1-x}$C$_x$)$_2$ single crystals (ref.~\onlinecite{Sologubenko05}), however, gave no evidence for a violation of the WFL in the same temperature region. In Fig.~\ref{LL0} we plot these earlier results together with the corresponding new data that we accumulated for Al-doped MgB$_2$. Below 1~K, the series of both C and Al doped materials exhibit ratios $L/L_0$ very close to 1. The smoothly increasing deviation from the ideal value with increasing temperature can easily be attributed to a non negligible $\kappa_{\rm ph}$ which, in the normal state of a metal exhibits a temperature variation that is typically stronger than linear in $T$.

Since it appeared that the WFL is well satisfied in MgB$_2$ with small amounts of defects due to Mg and B substitutions in the form of Al and C, respectively, it seemed of interest to verify whether the reported violation of the WFL is limited to pure MgB$_2$. We thus reexamined the electrical and the thermal transport properties of two new single crystals of MgB$_2$ (MB1 and MB2) between 0.5 and 6 K, both in the superconducting state in zero magnetic field and in the normal state, induced by a magnetic field of 50~kOe, oriented along the $c$ direction of the crystal lattice. The corresponding data resulted in the $L/L_0$ ratios that are also displayed in Fig.~\ref{LL0}.

In contrast to the substantial and temperature-dependent enhancement of the Lorenz number that resulted from the data that were earlier obtained for sample MB0, we now note an almost temperature-independent $L(T)$, somewhat smaller than $L_0$, for both new samples of pure MgB$_2$.  The reduction of $L(T)$ to below the value of $L_0$ is most likely not caused by inelastic scattering of the quasiparticles, because an almost parallel shift of $L(T)$ from $L_0$ would require that the responsible scattering would add a temperature-independent contribution to the thermal resistivity. However, all inelastic scattering processes that we are aware of, vary their strength with temperature. A more plausible explanation is to attribute the constant shift of $L(T)$ from $L_0$ to uncertainties in the sample geometry. If, as in our experiments, the same contacts are used to measure either the voltage or the temperature difference along the sample in the respective measurements of the electrical or the thermal conductivity, the same geometrical factor applies for the corresponding equations for calculating $\rho$ or $\kappa$. Thus in calculating $L(T)$, this factor drops out. This is really only true for cases, where the two contacts can be mounted far apart, in the sense that their separation $d$ is much larger than the contact width $\delta$. Our samples are quite small and hence $d \approx \delta$. In this case it is quite possible that the effective distances between the contacts for measuring the voltage or the temperature difference, i.e., $d_{\rho}$ or $d_{\kappa}$, are different. If now the ratio $d_{\kappa} / d_{\rho}$ differs from 1, the calculation of $L$ will result in a number that differs from $L_0$. For our small crystals, the distance between the contacts was of the order of 300 to 400~$\mu$m and the contact widths varied between 50 and 70~$\mu$m. It is therefore not unlikely that $d_{\kappa} \neq d_{\rho}$ which, in turn, may explain the constant shift of $L(T)$ from $L_0$.

In this sense we believe that no anomalous violation of the WFL is observed for the two new samples of single crystalline MgB$_2$. This suggests that our earlier observation reported in ref.~\onlinecite{Sologubenko02_KH} is not an intrinsic property of the material. We note, however, that the $\kappa(H)$ data exhibit the same features for all the investigated samples of pure MgB$_2$ and therefore, all the conclusions concerning the two-band features of MgB$_2$ drawn in ref.~\onlinecite{Sologubenko02_KH} remain valid.  

\section{Conclusions}

Based on experimental results of the electrical and thermal conductivity we note that the substitution of magnesium by aluminium in MgB$_2$ provokes an increase of the quasiparticle scattering of similar magnitude in both the $\sigma$- and the $\pi$ band of electronic states. The calculated reductions of the quasiparticle mean free paths in the $\sigma$- and $\pi$ band, respectively, are in agreement with results of de Haas-van Alphen experiments. The considerable violation of the Wiedemann-Franz law at low temperatures for pure MgB$_2$ that we reported in ref.~\onlinecite{Sologubenko02_KH} could not be confirmed in additional experiments using new single crystals of this compound. Since also for all the doped MgB$_2$ crystals that we studied since (ref.~\onlinecite{Sologubenko05} and this work), no anomalous deviations of the WFL were identified, our previous observation must have been of extrinsic origin. 

\acknowledgments
This work was financially supported in part by
the Schweizerische Nationalfonds zur F\"{o}rderung der Wissenschaftlichen
Forschung.

\end{document}